\documentstyle[12pt]{article}
\def\theequation{\thesection.\arabic{equation}}
\begin{document}

\def\beqra{\begin{eqnarray}} \def\eeqra{\end{eqnarray}}
\def\beqast{\begin{eqnarray*}}
\def\eeqast{\end{eqnarray*}} \def\beq{\begin{equation}}
\def\eeq{\end{equation}} \def\be{\begin{enumerate}}
\def\ee{\end{enumerate}}

%title page
\def\fnote#1#2{\begingroup\def\thefootnote{#1}\footnote{#2}\addtocounter
{footnote}{-1}\endgroup}

\def\ut#1#2{\hfill{UTTG-{#1}-{#2}}}

\def\sppt{Research supported in part by the Robert A. Welch Foundation and
NSF Grant PHY 9009850}

\def\utgp{\it Theory Group, Department of Physics \\ University of Texas,
Austin, Texas 78712}

\def\cO{{\cal O}}
\def\cj{{\cal J}}
\def\haf{\frac{1}{2}}
\def\lag{\langle}
\def\rag{\rangle}
\def\til{\tilde}

\def\gam{\gamma}
\def\Gam{\Gamma}
\def\la{\lambda}
\def\eps{\epsilon}
\def\La{\Lambda}
\def\si{\sigma}
\def\Si{\Sigma}
\def\al{\alpha}
\def\Tha{\Theta}
\def\tha{\theta}
\def\vphi{\varphi}
\def\del{\delta}
\def\Del{\Delta}
\def\ab{\alpha\beta}
\def\om{\omega}
\def\Om{\Omega}
\def\mn{\mu\nu}
\def\mun{^{\mu}{}_{\nu}}
\def\kap{\kappa}
\def\rsi{\rho\sigma}
\def\beal{\beta\alpha}

\def\rta{\rightarrow}
\def\eqv{\equiv}
\def\nab{\nabla}
\def\pa{\partial}

\def\ul{\underline}
\def\indt{\parindent2.5em}
\def\nd{\noindent}

\vspace*{24pt}
\parindent=2.5em

\ut{07}{95}

\vspace{6pt}
\hfill\today

\begin{center}

\large{\bf Collective Coordinates in String Theory}

\normalsize

\vspace{24pt}
Willy Fischler\fnote{*}{\sppt}, Sonia Paban$^*$, Moshe Rozali$^*$

\vspace{18pt}
\utgp

\vspace{18pt}

\end{center}

\abstract{The emergence of violations of conformal invariance in the
form of non-local operators in the two-dimensional action describing
solitons inevitably leads to the introduction of collective coordinates
as two dimensional ``wormhole parameters''.}

\vfill

\pagebreak
\baselineskip=21pt
\setcounter{page}{1}

\section{Introduction}

\indent\indent
This paper addresses the problems associated with describing scattering of
a spacetime soliton in
the context of string theory. More specifically we will present a way to
take into account the
changes in the soliton state due to the scattering. These may include
recoil as well as
changes in the various charges of the soliton. Our interest in this
question goes beyond the
scattering problem itself. Rather, we wish to address the conceptual puzzle
of how to properly
overlap different conformal field theories.

The organization of the paper is as follows:  We first review some general
properties of solitons
(for a complete discussion, including references, see \cite{one,two}). In
particular we show how
these scattering questions are resolved in pointlike theories. We explain
why the scattering off a
monopole is problematic in string theory. We then show how in string theory
the presence of
monopoles induces violations of conformal invariance in the form of bilocal
operators on the
worldsheet. One is led then to introduce  wormhole parameters that turn out
to account for the
finite number of degrees of freedom associated with the recoil or change of
charge.

For the sake of clarity we will focus on monopoles, but the conclusions
also apply to other
extended objects. Monopoles are time-independent, finite energy, stable
solutions of classical
field equations for weakly coupled systems. The weak coupling ensures that
these monopoles are heavy
and therefore can be reasonably well described by classical physics.

These solutions break in general some symmetries of the Lagrangian. For
example, the location in
space of the monopole breaks translational invariance: obviously, the
energy of the soliton is
independent of its center of mass location. This fact implies the existence
of normalizable ``zero
modes'' when quantizing the theory in a monopole background. These modes
are eigenfunctions of the
differential operator describing linearized fluctuations around the
monopole. The role of these
modes in pointlike theories is crucial in calculating the
scattering off a monopole when the monopole state changes. These are the
[Ascattering amplitudes we
focus on in this paper.

In pointlike theories such amplitudes are calculated using second
quantization (field theory). The
essential ingredient in the calculation is a change of variables that
trades each zero mode for a
collective coordinate. In the example of the recoiling monopole, there are
$D$ translational zero
modes that are exchanged for $D$ spatial coordinates of the monopole's
center of mass (in $D+1$
dimensions).

In transforming to these new coordinates care has to be exercised in
calculating the Jacobian of
the transformation. In a path integral formulation this involves a
straightforward but tedious
Fadeev-Popov procedure.

After having changed to these new variables, we are still dealing with a
degenerate perturbation
theory.  Indeed, monopoles located in different positions have the same
energy. It is important then
to use the right basis in the degenerate subspace in order to avoid
vanishing energy denominators.
(In diagrammatic language these energy denominators appear when there are
zero modes in intermediate
states, which then lead to spacetime infrared divergences). The correct
variables to use are the
components of the momentum conjugate to the center of mass position.

This summarizes the various steps that are necessary in pointlike theories
in order to calculate
scattering amplitudes.

In string theory this simple scattering presents some problems, since we do
not have a workable
second quantized formulation. In a first quantized string theory one
represents a heavy monopole
(positioned for example at the origin) by a two-dimensional conformal field
theory. In general
the monopoles initial and final states in a scattering process are
different. This  requires
interpolation between different conformal field theories, which in turn
would seem to require a
second quantized version of string theory.

However, as we show next, the recoil or change of charge of the monopole in
a scattering process
can be described in a first quantized formulation.

\section{Worldsheet wormholes}

\indent\indent
We will treat the monopole as a two dimensional conformal field theory
described by an action:
\beq
I=\int d^2z\sum_i\,g_i(x)\, \cO_i (\pa_\al x ,\cj_\al)\,. \eeq

$g_i(x)$ are functionals of the fields $x_\mu(z,z^*),\,\mu=0\ldots,D$ ,and
satisfy the
$\beta$-function equations:
\beq
\beta (g_i) = 0
\eeq

$\cO_i$ are functionals of $\pa_\al x$ and possibly other Kac-Moody
currents $\cj_\al$. Specific
actions describing monopoles and other solitons can be found for example in
\cite{thre}.

The proper treatment of the recoil of the soliton (change of conformal
field theory) emerges
through the existence of new violations of conformal invariance in the
conformal field theory (2.1).
These anomalies first appear at one loop and arise from a long thin handle
attached onto a sphere.
In the representation of higher genus surfaces as the plane with pairs of
discs cut out and
identified (Figure 1), the relevant limit in moduli space is shrinking one
pair of discs to zero
while keeping their distance fixed. The effects of such a handle in the
degeneration limit can be
represented by bilocal operators inserted on a surface of lower genus.

The existence of these world sheet divergences can be anticipated from the
spacetime interpretation. Indeed, these divergences are due to the
propagation in the degenerate handle of spacetime normalizable zero modes,
which are separated from the continuum by a gap. As mentioned above this is
related to vanishing energy denominators in a degenerate perturbation
theory, causing spacetime infrared divergences. In the worldsheet language
the divergences are manifest as worldsheet cut-off dependence, or in other
words conformal anomalies.

To regulate the violation of conformal invariance we introduce a short
distance cut-off $\epsilon$
on the worldsheet. To properly extract the worldsheet cut-off dependence we
use a formalism
developed by J. Polchinski \cite{for}. Given two closed Riemann surfaces
$\Sigma_1$ and $\Sigma_2$,
one can form a new surface $\Sigma$ by joining $\Sigma_1$ and $\Sigma_2$
with a long thin tube
(figure 2). One can then relate S-matrix elements on $\Sigma$ to those on
$\Sigma_1$ and $\Sigma_2$,
as given by the following expression:
\beqra
&&\!\!\!\!\!\!\! \int dm_{\Sigma} \left\lag \prod_i\int d^2
z_i\,V_i(z_i)\,\prod_k
B_k(z_k)\right\rag_\Sigma =
\nonumber \\
&& = \sum_a
\int\,\frac{dq}{q^{h_a-1}}\;\frac{d\bar{q}}{\bar{q}^{\tilde{h}_{a}-1}}\,
\int
d^2z_1\,\int d^2 z_2\int  dm_{\Sigma_1\oplus\Sigma_2} \\
&& \left\lag \prod_i \int d^2 z_i\;
V_i(z_i)\prod_k\hat{B}_k(z_k)\hat{\bar{b}}_{-1} \hat{b}_{-1}
\phi_a(z_1)\,\hat{\bar{b}}_1 \hat{b}_1\phi_a (z_2)\right\rag_{
\Sigma_1\oplus\Sigma_2} \nonumber
\eeqra
where $\phi_a$ are operators corresponding to a complete set of $L_0$ and
$\tilde{L}_0$
eigenstates
\beqra
L_0|\phi_a\rag &=& h_a|\phi_a\rag \nonumber \\
\tilde{L}_0|\phi_a\rag &=&  \tilde{h}_a|\phi_a\rag\;.
\eeqra

$q$ is the radius of the discs ($arg(q)$ is an identification angle). $\int
dm_\sigma$ is the
integration over the moduli of the respective surface $\sigma$ weighted by
appropriate ghost
insertions $B_k$ or $\hat B_k$ \cite{for}.

In the problem at hand the two surfaces $\Sigma_1$ and $\Sigma_2$ are the
same surface, and the
$q\rta 0$ limit corresponds to a long and thin wormhole attached to the
surface. It is clear from
(2.2) that possible logarithmic divergences (and hence cut-off dependence)
for small $q$ can  only
be generated by intermediate states satisfying $h_a=\tilde{h}_a=0$.
However, not all such states
contribute to a divergence. The zero momentum components
of the usual massless string states do not lead to divergences since they
are of zero measure in the
sum over states in (2.2). Therefore the only possible divergent
contribution can come from discrete
zero eigenstates of $L_0$ and $\tilde{L}_0$.

Such modes exist in the spectrum of $L_0$ and $\tilde{L}_0$. This is the
spectrum of linearized
fluctuations in the monopole's background. As discussed above, this
spectrum has zero eigenmodes.
These modes are normalizable in space and therefore are discrete. If
working in a finite (but large)
time interval $T$, they are a discrete part of the spectrum of $L_0$ and
$\tilde L_0$ (not just the
time independent part), thus enabling us to properly extract the divergent
contribution to (2.2).

For the sake of clarity we'll concentrate on the the zero modes associated
with translations in
space. The operators associated with these modes are:
\beq
\varphi_k=N\,c\bar c\;\frac{\pa g_i(x)}{\pa x}\;\cO_i(\pa_\al x,\,\cj_\al)
\eeq
where $N$ is a normalization constant. Using the Heisenberg equations of
motion one can write:
\beq
\varphi_\kappa =N\,c\bar c\,\pa_{\al}j^\al_\kappa
\eeq
where $j^\al_k$ are the two-dimensional N\"oether currents associated with
translations in space:
$$
x(z,z^*)\rta x(z,z^*)+q^\kappa
$$
The normalization constant is computed in the appendix and is found to be
\beq
 N^2=-\frac{4\pi}{\kappa}\,\log\epsilon\,.
\eeq

Where $\epsilon$ is the cut-off introduced earlier, and the constant
$\kappa$ is defined from the
following expectation value on the sphere;
\beq
\lag j^i(z)\,j^j (0)\rag=\frac{\kappa\delta^{ij}}{z^2}\,.
\eeq
The appearance of a divergent normalization constant is not surprising
since in the continuum
limit the operators $\pa_\al j^\al_i$ are BRST null and therefore have zero
norm.

Using the factorization formula (2.2), we see therefore that
the cut-off dependence due to a worldsheet wormhole can be summarized by the
following bilocal insertion:
$$
\sum_i\,\frac{1}{\kappa T}\,(4\pi\,\log\,\epsilon)^2\int d^2z_1\,\pa_\al
j_i^\al(z_1)\,\int d^2
z_2\,\pa_\al j_i^\al (z_2)
$$
where $T$ is the total (target) time introduced earlier.

The effect of a dilute gas of wormholes on the sphere exponentiates, and
can therefore be
summarized by a change in the action (2.1) on the sphere
\beq
\Delta I=\frac{1}{\kappa T}\,(4\pi\,\log\,\epsilon)^2\sum_i\int
d^2z_1\pa_\al j_i^\al(z_1)\;\int
d^2z_2\,\pa_\al j_i^\al(z_2)\,.
\eeq
The dilute gas approximation is justified in a weak string coupling limit.
As in the pointlike
case, we assume the system is weakly coupled.

\section{Scattering off the monopole}
\setcounter{equation}{0}

\indent\indent
We saw that in the dilute gas approximation the complete effect of the zero
modes (2.5) propagating  in the wormhole can be summarized by the inclusion
of a bilocal operator in the
world-sheet action. One can rewrite this bilocal as a local term in the
action by introducing
wormhole parameters that are integrated over \cite{fiv}:
\beq
\exp\left\{{\frac{1}{\kappa T}\left( 4\pi\,\log\,\epsilon\,\int d^2z\,\pa_\al
j^\al_i\right)^2}\right\} =
\int  \prod_i d\al_i
\,\exp\left\{{-\al_i^2-2\al_i\left(-\frac{4\pi\,\log\,\epsilon}{\sqrt{\kappa
T}}\right)\int d^2z\,\pa_\al j_i^\al}\right\}
\eeq
The wormhole parameters are like  two dimensional $\theta$ parameters (that
are integrated over),
since they multiply total derivatives.

As will be shown below, the wormhole parameters $\al^i$ are related to the
monopole's center of
mass momentum. In other words one can interpret \break $\exp\{\,{-\al^i}
\,N\int d^2z\,\pa_\al
j^\al_i(z)\}$ as a ``vertex operator'' for the center of mass of the monopole
with momentum
 proportional to $\vec\al$. In order to see this, consider the example of a
matrix element
corresponding to the elastic scattering of a string quantum off the
monopole. The matrix element can
be expressed as

\beqra
\!\!  \lag V_1V_2\rag &=& \int \prod_i d\al_i e^{-\al_i^2} \int Dx
\exp\left\{{-\int d^2z\sum_j
g_j\cO_j}\right\}\,\exp\left\{{\frac{8\pi\al^i\log\epsilon}{\sqrt{\kappa
T}}\int d^2 z\pa_a
j^\al_i}\right\} \nonumber\\
 && \hspace{2in} \int d^2z_1 V_1(z_1)\,\int d^2z_2 V_2(z_2)
\eeqra

The expression is evaluated on the sphere (to leading order). $V_1$ and
$V_2$ are vertex operators,
which can be written as $h(x)\cO(\pa_\al x,\cj_{\al})$, where $h$ are
solutions of the linearized
$\beta$-function equations. For large value of $\vec x$ these vertex
operators have the asymptotic
behaviour
$$ h(\vec x,t)\sim \,e^{i\vec k\cdot \vec x+i\delta_k}\,e^{ -i\omega t} $$
where $\vec k$ is a spatial momentum and $\delta_k$ is a phase shift. This
is identical to the
pointlike case; asymptotically (for large values of $\vec x$) the spatial
momentum becomes
a good quantum number,
reflecting the fact that the
monopole energy is concentrated in a finite region of space.

The scattering amplitude in (3.2) is computed with a new term in the
action, proportional to the
wormhole parameter $\vec \al$. The effects of this term can be easily
evaluated. First, one needs to
normal order this new term. This is easily shown to yield (Using results
from the appendix) :
\beq
\exp\left\{{\frac{8\pi \al^i\,\log\,\epsilon}{\sqrt{\kappa T}}\int d^2z\pa_\al
j^\al_i(z)}\right\}
=\exp\left\{{-\frac{8
\pi\al_i^2\,\log\,
\epsilon}{T}}\right\}:\;\exp\left\{{\frac{8\pi\al^i\,\log\,\epsilon }
{\sqrt{\kappa T}}\,\int d^2z\,\pa_\al j^\al_i(z)}\right\}:
\eeq

Also, since $j_i^\al$ are the N\"oether currents associated with spatial
translations,
$\exp\left\{{\int d^2z\,\pa_\al j^\al_i(z)}\right\}$ acts as a translation
operator (This is easily
verified using the Ward identity associated with translation). Therefore
\beq
\exp\left\{{\frac{8\pi \al^i\,\log\,\epsilon}{\sqrt{\kappa T}}\,\int
d^2z\pa_\al j^\al_i}\right\}
\prod_i V_i(x)=
\exp\left\{{\frac{8\pi\al^i\log\,\epsilon}{\sqrt{\kappa T}}\,\frac{\pa}{\pa
x}}\right\} \prod_i
V_i(x)
\eeq

Finally, one has to contract the vertex operators among themselves. This
can be done using the
operator formalism on the sphere,
$$
\lag V_1V_2\rag=\lag V_1|V_2\rag=\sum_a \;\lag V_1|\phi_a\rag\lag\phi_a|V_2\rag
$$
The states $|\phi_a\rag$ are defined in (2.3). The logarithmic dependence
on the cut-off arises from
the states annihilated by $L_0$ and $\tilde L_0$, as discussed above.
Therefore:
$$
\lag V_1V_2\rag=N^2\lag V_1|\pa_\al \,j^\al_i\rag\lag\pa_\al j^\al_i |
V_2\rag + {\rm finite~ terms}
$$
using the Ward identity mentioned above we obtain:
\beq
V_1V_2=- \sum_i\,\frac{4\pi\,\log\,\epsilon}{\kappa}\;\frac{\pa V_1}{\pa x_i}\;
\frac{\pa V_2}{\pa
x_i}\,.
\eeq
The above results enable us to write the matrix element (3.2) as follows:
\beqra \lag V_1V_2\rag&=&
\int \prod_i d\al_i e^{-\al_i^2} \exp\left\{{-\frac{\pi}{2}\,\log\,\epsilon
\left(
\frac{4\al_i}{\sqrt{T}}-\frac{2}{\sqrt{\kappa}}\;\frac{\pa}{\pa
x}\right)^2}\right\} \nonumber \\
&& ~~~~~~~~~~~~~~~\int Dx\,e^{-I} \int d^2z_1 V_1(z_1)\,\int d^2z_2
V_2(z_2)\,. \eeqra

Imposing conformal invariance and going to the continuum limit then forces
the wormhole parameter to
satisfy
$$\vec \al=\sqrt{\frac{T}{4\kappa}}\;(\vec k_1+\vec k_2)\,. $$
This relation is enforced through a
$\delta$-function, $\del^{(D)}\left(\vec\al-\sqrt{\frac{T}{4\kappa}}(\vec
k_1 +\vec k_2)\right)$.
Therefore conformal invariance forces the monopole to recoil with the exact
momentum needed to
conserve the total momentum.

Performing the  integration over $\vec \al$ results in the S-matrix element
acquiring an additional
phase shift
{}~~$\exp\,{-\frac{(\vec k_1+\vec k_2)^2 T}{4\kappa}}$ ~~due to the recoiling
monopole. By Lorentz
invariance in spacetime, the two-dimensional quantity $\kappa$ (calculated
on the sphere), is to be
identified with half the classical mass of the monopole. We conjecture this
relation between
$\kappa$ and the mass to hold for an arbitrary solitonic background.

One also expects the energy conservation condition to be modified to
include the recoil of the
monopole. We expect in this respect a violation of conformal invariance for
the S-matrix element of
the form
$$
\exp\left\{{-\log\,\epsilon\left( \omega_1+\omega_2-\frac{(\vec k_1+\vec
k_2)^2}{4\kappa}\right)^2}\right\}\,.
$$

Conformal invariance will then be restored by requiring the modified energy
conservation condition.
However, the modification due to the recoil will only appear in two loop
order. Indeed, the term~~
 $\log \epsilon\,\frac{(\vec k_1+\vec k_2)^4}{16\kappa^2}$ ~~appearing in
the above expression can
only be obtained through the dependence of  the matrix elements on the
wormhole parameters
$\al^i$ raised to the fourth power.

\section{Conclusions}

\indent\indent We have seen that in the presence of solitons the two
dimensional field theory
displays non-local violations of conformal invariance. The wormhole
parameters that are then
introduced appear as two-dimensional $\theta$ parameters that are
integrated over. There is one
such parameter for each quantum mechanical degree of freedom associated to
the recoil of the
soliton. Requiring the theory to be ultimately conformally invariant, fixes
the values of these
wormhole parameters such that the soliton recoils with overall momentum
conservation. Obviously all
that  has been said about recoil applies to charge deposition or any other
change in the quantum
numbers of the monopole. Similar non-localities will appear on the
worldsheet in the case of
space-time instantons because of the existence of normalizable zero modes.

A more challenging
situation occurs when describing multimonopole scattering. As was reviewed
by Atiyah and Hitchin in
the BPS limit
\cite{sez} this scattering can be described by the motion of collective
coordinates on an
hyperk\"ahler manifold. It would be interesting to understand how in this
case the interaction
between wormhole parameters appear from degenerating surfaces of higher
genus such as to reproduce
the known dynamics of the collective coordinates.

More generally the emergence of such non-local violations of conformal
invariance will lead to the
proper measure for summing over two-dimensional field theories which may be
required in certain
circumstances in string theory, like in the example discussed above.

\section{Acknowledgements}
\indent\indent
We would like to acknowledge useful conversations with J. Russo at an
earlier stage of this work.
We would like also to thank V. Kaplunovsky, M. Dine, J. Polchinski and L.
Susskind for discussions.
This work is supported in part by the Robert A. Welch Foundation and NSF
Grant PHY 9009850.

\section*{Appendix}
\indent\indent
The proper way to normalize states appearing in (2.2) was discussed in
\cite{for}. This leads to
the following normalization condition for the operators (2.5) :

\renewcommand{\theequation}{A.\arabic{equation}}
\setcounter{equation}{0}
\beq
N^2 \int\limits_{S^2} \,DxDcDb\, e^{-I}\,
c_0(0)\bar{c}_0(0)\,\varphi_i(0)\varphi_j(z)=4\pi
\delta_{ij}
\eeq

The point $z$ is arbitrary since in the special case of the operators (2.5)
the expression is
conformally invariant. Performing this functional integration gives: \beq
-\haf \,N^2\kappa\,|z|^4\left(\bar\pa^2\,\frac{1}{z^2}+c\cdot c\right)=4\pi
\eeq
where we have used:
\beq
\int\limits_{S^2}\, Dx\, e^{-I}  \,j_i(z)\,j_j(0) = \lag
j_i(z)j_j(0)\rag=\frac{\kappa\del_{ij}}{z^2}\,.
\eeq

This integral is determined by it's conformal weight, up to the arbitrary
constant $\kappa$.

The expression (A.2) is independent of the choice of $z$, so we can
integrate over $z$ (with the
conformally invariant weight). This integration gives:
$$ 4\pi\,N^2 \kappa=4\pi\,\int\frac{d^2z}
{|z|^2} =-16\pi^2\log \epsilon + {\rm finite} $$
or
\beq
N^2=-\frac{4\pi\log\epsilon}{\kappa} + {\rm finite}\,.
\eeq
\pagebreak

\section*{Figure Captions}

\begin{quote}
Figure 1: Example of a surface with one handle.

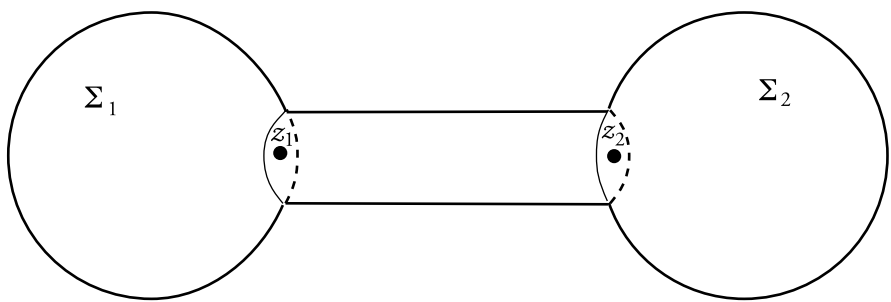
Figure 2: The surface $\Sigma$.
\end{quote}

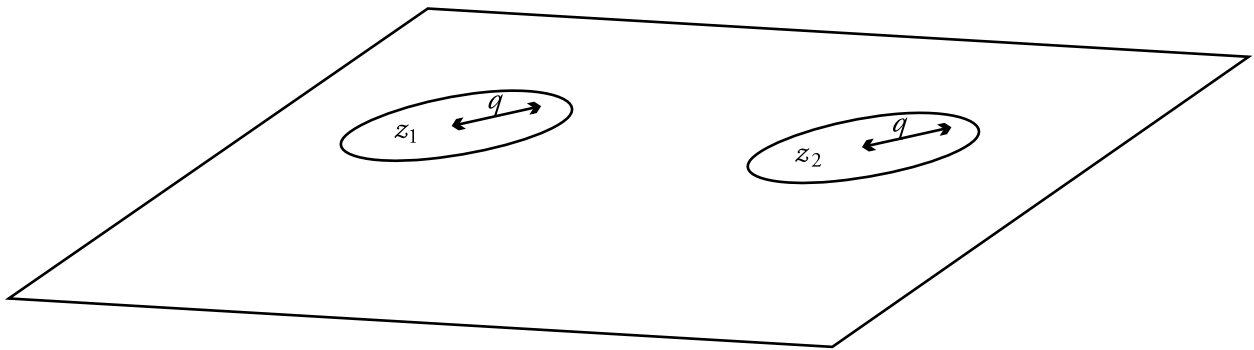
\end{document}